\documentclass{elsarticle}

\usepackage{hyperref,comment,color,float}

\journal{APS}

\begin{document}

\begin{frontmatter}

\title{
Sector-wise analysis of Indian stock market: Long and short-term risk and stability analysis}

\author{Suchetana Sadhukhan\fnref{myfootnote1}}
\address{Department of Physics, School of Advanced Sciences and Languages, VIT Bhopal University, Kothri Kalan, Sehore-466114, India}
\fntext[myfootnote]{Corresponding author}

\author{Poulomi Sadhukhan\fnref{myfootnote}}
\address{Bennett University, Plot Nos 8-11, TechZone II, Greater Noida, Uttar Pradesh - 201310}

\begin{abstract}
This paper, for the first time, focuses on the sector-wise analysis of a stock market through multifractal analysis. We have considered Bombay Stock Exchange, India, and identified two time scales, short ($<200$ days) and long time-scale ($>200$ days) for investment. We infer that long-term investment will be more profitable. For long time scale, sectors can be separated into two categories based on the Hurst exponent values; one corresponds to stable sectors with small fluctuations, and the other with dominance of large fluctuations leading to possible downturns in those sectors.
\end{abstract}

\end{frontmatter}

\section{Introduction\label{sec:intro}}
Multifractal behavior can be perceived in various complex systems which also includes ﬁnancial market. The study of multifractalilty of the time-series of stock prices of financial market has gained a lot of interest recently in order to understand the market behaviour. The multifractal behavior is identified from the presence of multiple Hurst exponents which are extracted from time series data by detrending the fluctuations. The presence of multifractal behavior has been studied and established  as robust behavior of the financial time-series for wide variety of data with different frequencies viz, intraday closures, high frequency prices \cite{aslam2020, aslam2021, zhang2021} etc. and for markets in different countries \cite{aslam2021,milocs2020}. There are several existing methods of studying multifractality with different strengths and weaknesses, most common of which are Multifractal Detrended Fluctuation Analysis (MFDFA), Multi-ractal Detrended Moving Average (MFDMA), Multi-fractal detrended cross-correlation analysis (MFDXA) etc\cite{kantel02, kantel03, he17, sim19, zho12, sus20, charutha2020, lai19}. The multifractality of financial time-series has been tested for developed  and developing countries \cite{milocs2020,arshad19,  gu2013, dutta2010, hasan15, caraiani2012evidence, grech2009, tiwari2019} by comparing the degree of multifractality. 
As the internal structure of the market is responsible for defining the state of the market, the multifractal nature of financial time series also provides an alternative way of studying the market stability and risks, correlations in stock prices \cite{gu2013, rizvi14}, and signature of crashes \cite{dutta2010, hasan15, anag16}, etc.

In this paper, we study sectoral indices of Bombay Stock Exchange (S\&P BSE), India. 
The multifractal behavior of overall Indian stock market has been studied by S Dutta, Kumar et. al., Nargunam et. al \cite{dutta2010, kumar2009, nargunam2022} and others. Multifractality being an inherent nature of any financial time series data, Indian financial market also does not deviate from it. 
To understand the behavior of stock market in detail, we look into the underlying structure within the Indian stock market. Instead of looking at the overall market behaviour, in this paper we have done a sector-wise study, which has not been done for any financial market till now. Though the overall Indian market has been categorized as a persistent developing market \cite{kumar2009, nargunam2022}, we clearly can see two separate categories of sectors in terms of stability and risk assessment. 

In Sec. \ref{sec:data}, the detailed description of the financial time-series data has been given with the sectors considered for analysis. The algorithm of modified MFDFA has been used along with the overlapping moving window (OMW) and ensemble empirical mode decomposition (EEMD) \cite{lai19,zhou10, qian11, wu09,  lee18}. OMW-EEMD-MFDFA method is used for analysis has been described in Sec. \ref{sec:method}. The results are discussed in Sec. \ref{sec:res} and we conclude at Sec. \ref{sec:conc} with the anticipation that this study will help the investors to make the strategy to choose right sectors and investment duration for higher chance of profits.
\section{Data Description}\label{sec:data}
For the time series analysis, we have collected daily closure prices of all the sectors' indices of Bombay Stock Exchange (S\&P BSE) from 2nd January 2017 to 30th June 2022 ($1362$ trading days) from the BSE India website \cite{bse}. We consider only the sectors, $22$ in total, for which data are available for the full period in the given date range. We do not have any missing data points in the considered data-set. Detailed information about the sectors is given in table \ref{table:stocksdetails1}. 
\begin{table}[ht]
  \centering
  \resizebox{0.65\textwidth}{!}{
  \begin{tabular}{|c|c|c|c|c|}
    \hline
    \cline{1-3}
    \textbf{Symbol} & \textbf{Sector} & \textbf{$\#$ of stocks}\\
    \hline
    AU & Auto & 15 \\ \hline
    BM & Basic Materials & 189 \\ \hline
    BX & Bankex & 10 \\ \hline
    CD & Consumer Durables & 12 \\ \hline
    CDGS & Consumer Discretionary Goods \& Services & 297 \\ \hline
    CG & Capital Goods & 25 \\ \hline
    CPSE & CPSE & 52 \\ \hline
    EG & Energy & 27 \\ \hline
    FMCG & Fast Moving Consumer Goods & 81 \\ \hline
    FN & Financials & 139 \\ \hline
    HC & Healthcare & 96 \\ \hline
    ID & Industrials & 203 \\ \hline
    II & India Infrastructure & 30 \\ \hline
    IT & Information Technology & 62 \\ \hline
    MT & Metal & 10 \\ \hline
    ONG & Oil \& Gas & 10 \\ \hline
    PSU & PSU & 56 \\ \hline
    PWR & Power & 11 \\ \hline
    RE & Realty & 10 \\ \hline
    TC & Telecom & 17 \\ \hline
    Teck & Teck & 28 \\ \hline
    UT & Utilities & 24 \\ \hline
     \end{tabular}
}
  \caption{Details of Bombay stock exchange (S\&P BSE) sectors considered for the analysis are given here. The first column consists of the symbol of the sectors followed by the name of the corresponding sectors in the second column.  The third column shows the total number of available stocks in each sector during the time period considered for analysis. For the rest of the paper, we are going to refer the sectors by their corresponding symbol name provided here.}
  \label{table:stocksdetails1}
\end{table}
\section{Methodology}\label{sec:method}
\subsection{Logarithmic Return}
We compute the logarithmic return of sector indices' daily closure values with $\Delta t= 1$ day using the formula,
\begin{equation}
r_i(t)=\ln \; P_i (t+\Delta t)- \ln \; P_i(t), \qquad i=1, 2, ...,k.
\end{equation}
Here $r_i(t)$ represents the logarithmic return price of $i$-th sector at day $t$ with $t = 1, 2,...,T-\Delta t$. $T$ and $k$ are the total number of trading days and sectors available in the considered time period, respectively. 
\subsection{EEMD and OMW based Multifractal Detrended Fluctuation Algorithm (EEMD-OMW-MFDFA)}
Multifractal Detrended Fluctuation Algorithm (MFDFA) is a well-accepted and a hugely used method to find the presence of self-similarity and fractionality in any stochastic time series with different power scaling laws \cite{kantel02, gan15, krzy19, adarsh19, bara19}. It can be used as an indicator to analyze the dynamics of data-set quantitatively as well as qualitatively.
The model was first developed as a Fluctuation Analysis. But this method was incompatible in dealing with the time-series data with a long-term trend expected in real-world data analysis. Peng et al. \cite{peng94} developed Detrended Fluctuation Analysis (DFA), where the data-set is detrended through polynomial fitting to consider the non-stationarity of the data, which in later years modified to a continuum of power variations of DFA, i.e., MFDFA by Kandelhardt et al. \cite{kantel02, kantel03, mandel97}. DFA becomes a particular case of MFDFA with a specific choice of power parameter (power $q=2$). Instead of using polynomial fitting to detrend the data, we have used ensemble empirical mode decomposition(EEMD) for the same \cite{zhou10, qian11, wu09}. As our data-set is not very long, we have considered overlapped moving windows (OMW), for a better analysis. The overall modified algorithm of EEMD-OMW-MFDFA consists of five steps as follows \cite{mandel97, movahed06, ihlen12};
\begin{itemize}
    \item Normalize the data-set by subtracting the average value from each data points and computing the cumulative sum
\begin{equation}
y(N) = \sum_{k=1}^N [x(k)-\overline x]
\end{equation}
where $x(k)$ is the time series of length $N$ at any point $k$ and 
$\overline x$ 
denotes the average over the entire data-set.
\item As the second step, we consider a moving window of segment size $s$ shifted by 1 data point as overlapped moving window (OMW) \cite{ lai19,zhou10, lee18} and replace the conventional equally divided non-overlapping segmentation to overcome the issue of poor statistics due to lesser number of segments for larger lags $s$. Hence, total number of epoch will be $N_s=N-s+1$ for time-series of length $N$.
\item In the next step, we replace the local polynomial fitting for detrending the time series \cite{qian11} by EMD method \cite{wu07}. Here, the data trends are computed by removing the Intrinsic Mode Functions (IMFs) obtained by EMD method to get stationary data \cite{zhang19}.
Conditions and steps included in the calculation of intrinsic mode functions are as follows \cite{liu20}.

Conditions:
\begin{enumerate}
 \item Difference between the number of extremes and zero crossings should be one, at maximum.
 \item Local mean value of the envelope at any point, defined by the local extrema, must be zero. To get the best-fitted curve between the local extremes, the cubic splines method is considered with the above condition.
\end{enumerate}
Steps:
\begin{enumerate}
 \item Find an IMF which satisfies the conditions. as mentioned above.
 \item Compute the remaining time series by subtracting the IMF from the original time series.
 \item Repeat the (i) and (ii) steps until one has the final IMF as a purely residual trend.
\end{enumerate}
A sifting process is considered for the decomposition with the detailed steps as follows \cite{liu20}:
\begin{enumerate}
 \item Figure out all the extrema of the original time series $x(t)$ and use the maxima $U(x)$ and minima $L(x)$  to interpolate an upper and lower envelope, respectively.
 \item Calculate the mean $\mu(t)$ of the envelope: $\mu(t)= \frac{U(x)+L(x)}{2}$ and subtract it from the original time series: $g(t) = x(t) - \mu(t)$.
 \item If the subtracted time series g(t) satisfies the conditions mentioned above for IMF, then the sifting process stops. If not, then the procedures from (i) to (iii) are repeated with the subtracted time series $g(t)$ in place of $x(t)$.
 \item Once the sifting process stops, say after the $n$-th steps, we subtract the final IMF to obtain the pure trend of the data-set $r_n(t)$: $r_n(t) = x(t) -\sum_{i=1}^n g_i(t)$ \cite{liu20}.
\end{enumerate}
A disadvantage of considering EMD is mode mixing \cite{wu09, liu20}. IMF may consist of different frequency components. On the other hand, the same frequency or frequencies of similar magnitudes can be found in different IMFs. To overcome this, we consider the ensemble EMD (EEMD) \cite{wu09,huang08}. For the modification, we now add different white noises $(w_i(t))$ to original data set $(x(t))$: $x_i(t)=x(t)+ w_i(t)$ where $i=1, ..., M$ for ensemble size $M$ and compute the IMFs using EMD method which satisfies all the conditions. Finally, we consider ensemble mean of all the IMFs to get the final one, and we calculate the trend by subtracting the ensemble-averaged IMF from the original time series.\\
To choose the white noise we have considered: $\epsilon_n=\frac{\epsilon}{\sqrt N}$
with $\epsilon_n$, the final standard deviation of the error, is calculated from the difference between the
original time series data and the corresponding IMFs, and $\epsilon$ is the amplitude of the added noise. We have considered the ensemble number as $100$ and the amplitude of added white noise as $0.2$ times the data set's standard deviation \cite{movahed06,zhang19, liu20,  agbazo21}.\\
Another disadvantage of the EMD method is it leads to numerical errors, which results in incorrect decomposition analysis. To discard the spurious IMF, we
include an extra threshold condition based on correlation with the
original signal, which will be significant for relevant IMFs \cite{agbazo21,peng05, ayenu10}. We only consider the
IMFs which are having correlations larger than the threshold correlation value. The threshold correlation value $\mu_{TH}$ is:
\begin{eqnarray}
\mu_{TH}=\frac{max(\mu_i)}{10 \times max(\mu_i)-3} \qquad i=1, 2, ...,k
\end{eqnarray}
Here $\mu_i$ is the correlation coefficients between $i$-th IMF and the original data set and $k$ is the total number of IMFs.
\item Finally, the $q$-th order fluctuation function $F_q(s)$ is computed as follows considering all segments $s$:
\begin{equation}
F_q(s)=\left[ \frac{1}{2N_s}  \sum_{v=1}^{2N_s} [F^2(s,v)]^{q/2} \right]^{1/q} \qquad \qquad for \; q\ne0
\end{equation}
\begin{equation}
F_0(s)=exp \left[  {\frac{1}{2N_s}\sum_{v=1}^{2N_s} ln [F^2(s,v)]^{1/2}}  \right]  \qquad for \; q=0
\end{equation}
according to L’H\^ospital’s rule.
\item The fluctuation function $F_q(s)$ varies as power law of the segment size $s$ with a function of power $q$ for a multifractal time-series \cite{hurst51},
\begin{equation}
F_q(s) \sim s^{H(q)}
\label{12}
\end{equation}
Here function $H(q)$ is defined as the generalized Hurst index or self-similarity exponent of the multifractal data-set. For a mono-fractal time-series scaling behavior is independent of power $q$ and shows the same behavior for small $(q < 0)$ and large $(q> 0)$ fluctuations.
\end{itemize}
\subsection{Multifractal Strength Analysis}
To analyze the the strength of multifractality in the considered time-series, we compute multifractal scaling exponent $\tau(q)$ using generalized Hurst exponent calculated from eqn. \ref{12} by 
\begin{eqnarray}
\tau(q) = qH(q) -1
\label{13}
\end{eqnarray}
This can also provide the information about the shape of the singularity spectrum through Legendre Transformation \cite{hentschel83, halsey86, kurths87, meneveau87}. Singularity index or H$\ddot{o}$lder exponent $\alpha$ can be computed by  differentiating a smooth $\tau(q)$ function: $\alpha = \tau'(q) = H(q) + qH'(q)$ and singularity spectrum is related to the singularity strength as $f(\alpha) = q \alpha - \tau(q)$ \cite{salat17}. The width of the inverted parabolic singularity spectrum curve ($\Delta \alpha$) refers to the strength of the multifractality and complexity of the time-series: $\Delta \alpha =\alpha_{max} - \alpha_{min}$ with $\alpha_{max}$ and $\alpha_{min}$ as the maximum and minimum value of the $\alpha$. In other words, it also denotes the uneven distribution and presence of more severe fluctuation in the data-set. 
\section{Results and Analysis}\label{sec:res}
For the fluctuation analysis, we have log normalized the return data of the daily closing prices of each stock. We show the same for Au sector, in Fig. 1 (bottom), compared to the return data set (up). All the other sectors also show similar structure. 
\begin{figure}[ht]
	\centering
	\includegraphics[width=10cm]{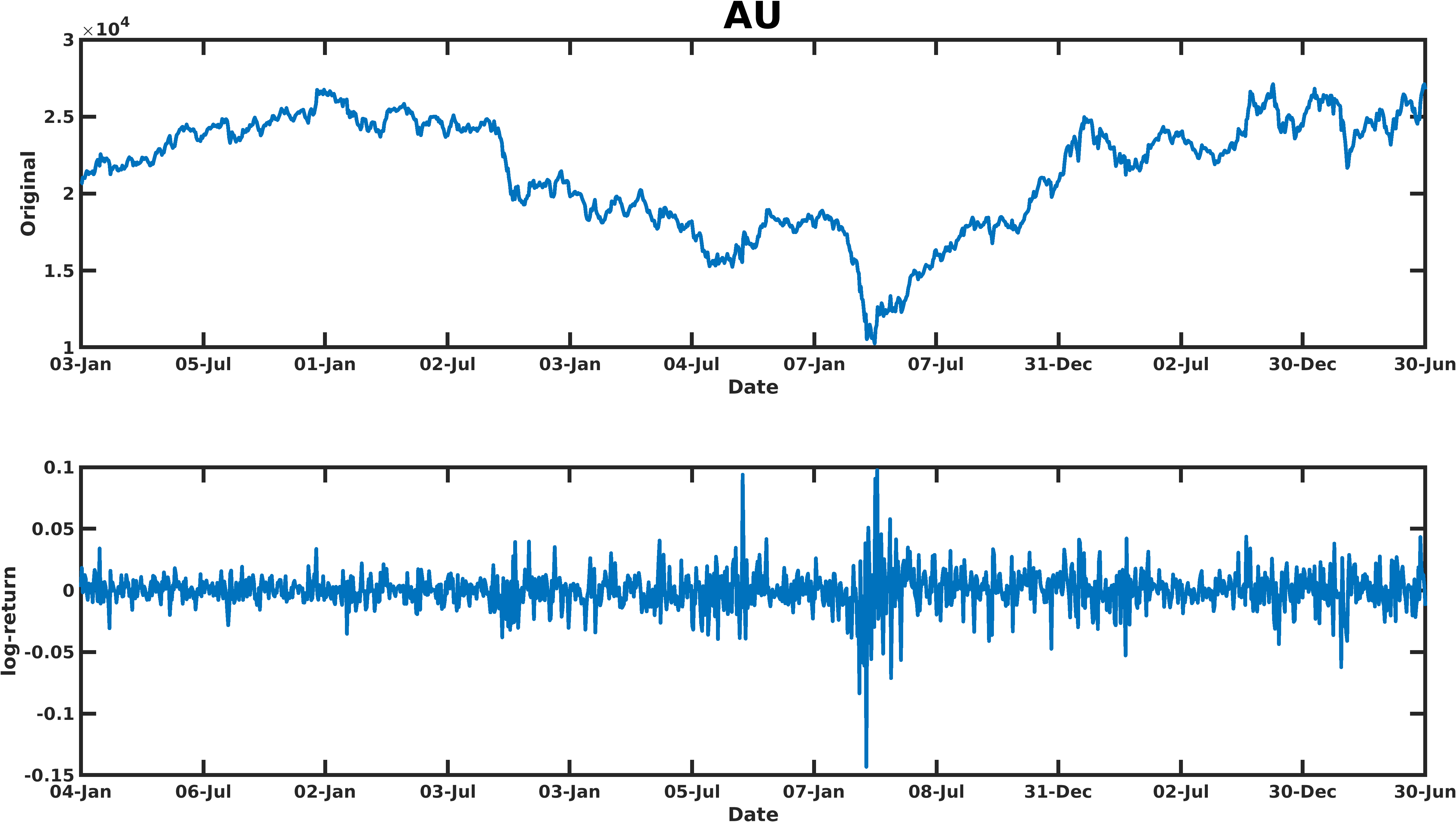}
\caption{Daily closing prices (top) and normalized log-returns (bottom) of the full time period for Auto sector is shown here.}
\end{figure}
Earlier the presence of two scaling behavior (periods shorter and
longer than 24 hours) has been found for price time series \cite{weron04, simonsen03, gorjao21, gorjao22, han22}. In our analysis of sectorwise analysis,  we similarly found two separate time-scales with a shorter timescale between $10$ days to $200$ working
days, and a longer time scale between $200$ and $1000$ working days. We show the shorter time scale plots in Fig \ref{short}. 
\begin{figure}[ht]
	\centering
	\includegraphics[width=10cm]{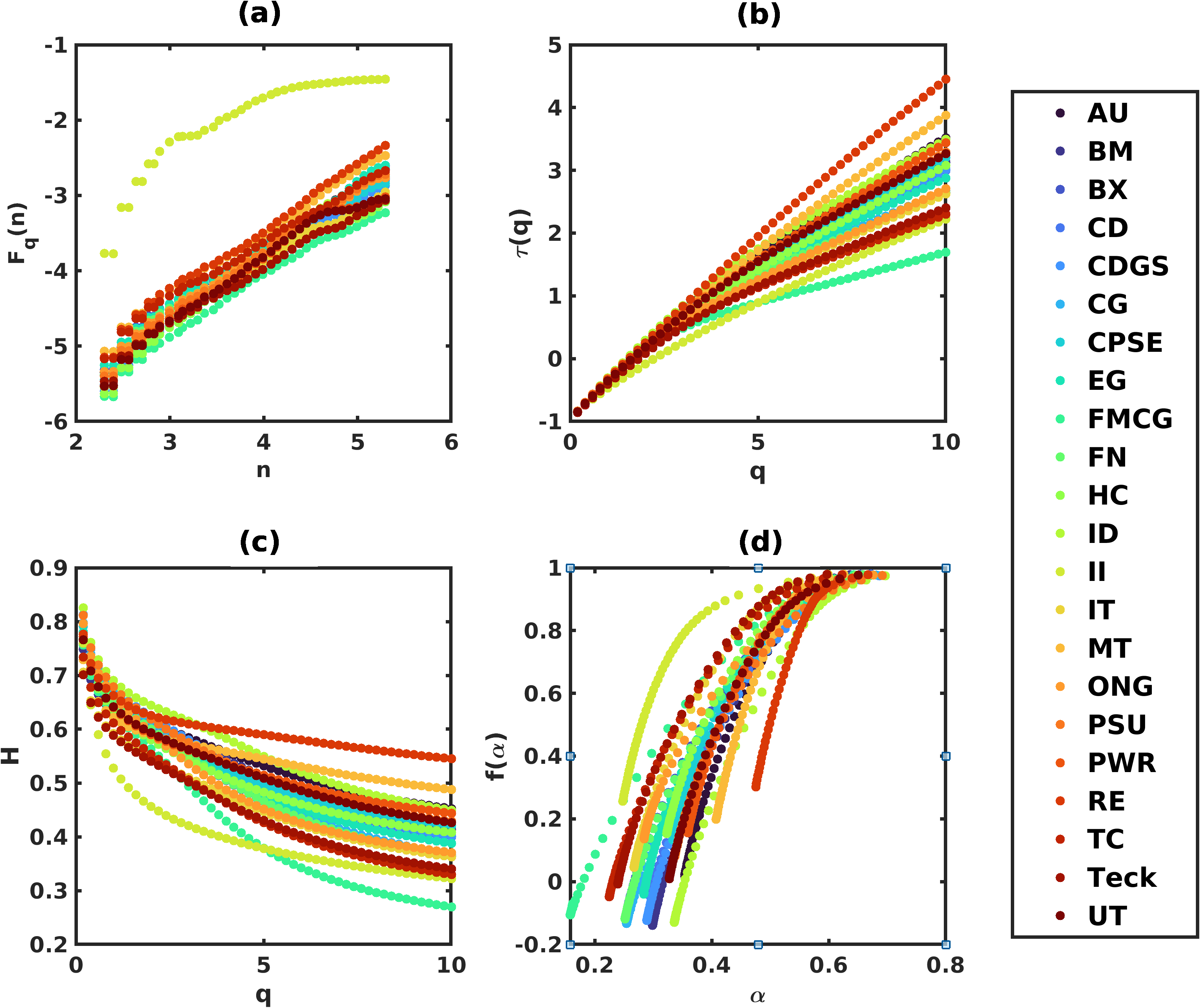}
\caption{Multifractal analysis for Short term (upto $200$ days) period. (a) Fluctuation functions for various sectors (see color code), 
(b) Generalized Hurst exponent depending on $q$,
 (c) Mass exponent $\tau(q)$ vs $q$, 
 (d) Multifractal spectrum.}
 \label{short}
\end{figure}
Panel (a) shows the fluctuation function $F_q(s)$ for $q=2$. As the graph shows, in the log-log scale, it possibly fits a straight line and thus indicates a presence of scaling with $q$. One can notice the non-existence of negative $q$ power in Fig \ref{short}, (b) $\tau(q)$ vs $q$ and (c) $H(q)$ vs $q$, which was also reported in an earlier paper \cite{gorjao22}. On the other hand, for long-time scale, the whole range of $q$, both negative and positive values, are present (see Fig. \ref{long}). As for the shorter segments $s$, the average variance is not well defined, which leads to the non-existence of a negative power value of those variances. But, on advantage, it establishes a threshold value to mark the change in the multifractal behavior in the market. 
\begin{figure}[ht]
	\centering
	\includegraphics[width=10cm]{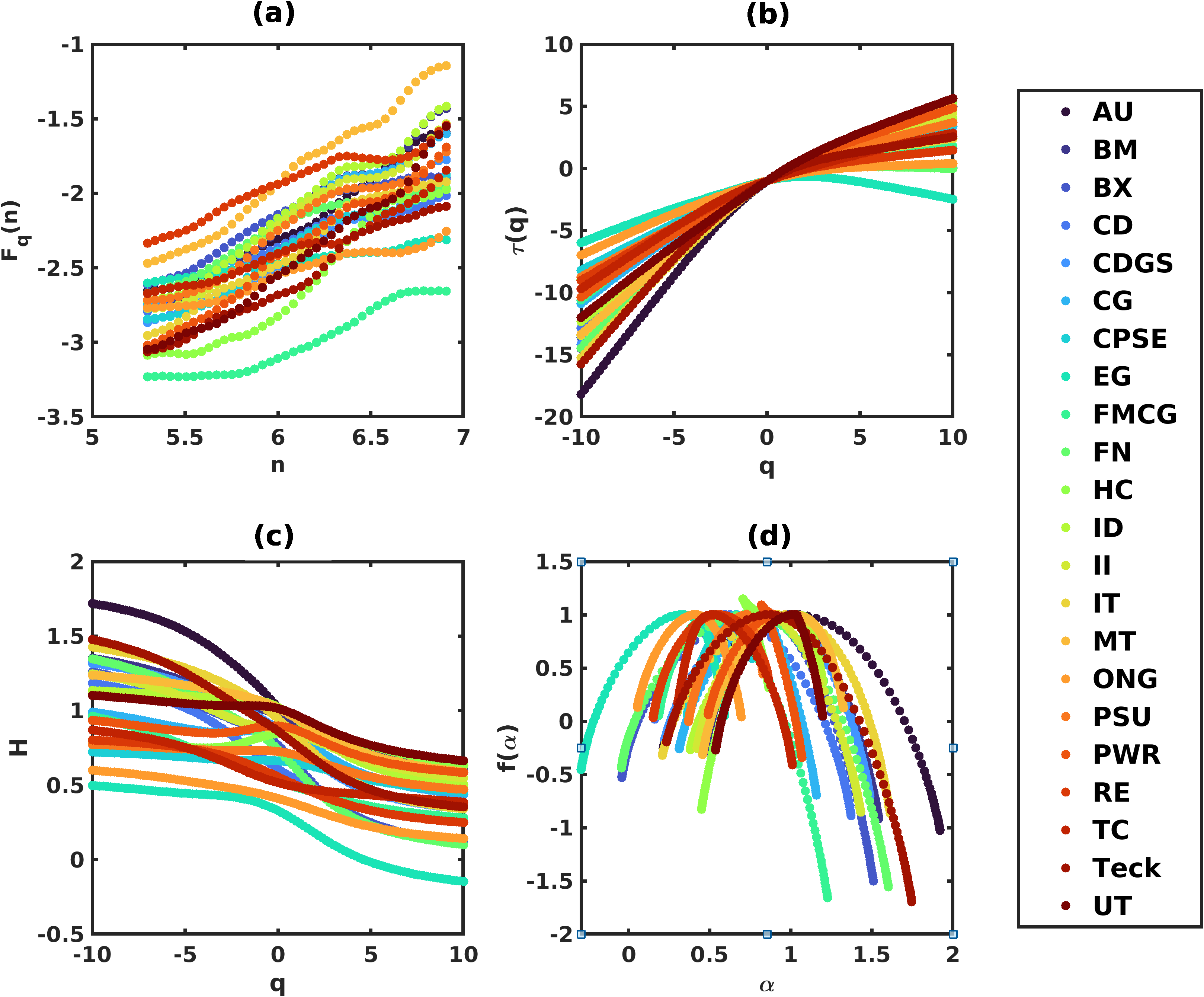}
\caption{Multifractal analysis for long term period (more than $200$ days). (a) Fluctuation functions for various sectors (see color code), 
(b) Generalized Hurst exponent depending on $q$,
 (c) Mass exponent $\tau(q)$ vs $q$, 
 (d) Multifractal spectrum.}
 \label{long}
\end{figure}
Scaling exponent from the Fluctuation function at $q = 2$ also helps to calculate the Hurst exponent $(H_2)$ for stationary time series from eqn. \ref{13}. The generalized Hurst exponent is dependent on $q$ and decreases as $q$ increases, which is a characteristic of multifractal features of the data set \cite{zunino08}. Renyi exponent $\tau(q)$ is also non-linear, indicating multifractality. Finally, we obtain the single-humped multifractal spectrum $f(\alpha)$ ($\alpha$=singularity strength) by the Legendre transform, indicating multifractality. The width of the multifractal spectrum denotes the power or degree of multifractality. We also calculate the asymmetry of the spectrum by the ratio $B = (\delta\alpha_L -\delta \alpha_R ) / (\delta\alpha_L +\delta \alpha_R )$ where $\delta\alpha_L=\alpha_0-\alpha_{min}$ and $\delta\alpha_R=\alpha_{max}-\alpha_0$ are the width of left and right branches of the singularity spectrum. These two parameters decides the pattern of high and low fluctuations \cite{agbazo21,agter01}.  $\alpha_0$, $\alpha_{max}$, $\alpha_{min}$ are the values of $\alpha$ at maximum value of $f(\alpha)$, maximum and minimum $\alpha$. The value of $B$ ranges from $-1$ to $1$ \cite{hou18, xie04}. It denotes the skewness of the multifractal spectrum. $B=0, >0, <0$ signifies the symmetric (low and high fluctuation are almost equally responsible for the dynamics), left-skewed (large fluctuation is more dominant), and right skewed (dominance of low fluctuation is more clearly visible) multifractal spectrum, respectively. For the shorter time scale, the multifractal spectrum is vast, which signifies very strong multifractality or high volatility and the occasional burst of the stock market. One can conclude that this shorter period is not suitable for the investment point of view. In contrast, we have the full singularity spectrum for large time scales, and less width denotes a more stable market suitable for investment for more extended time scales. 
Table \ref{table2} shows the most important parameters of the multifractal spectrum: $\Delta \alpha$, $H_2$, $dH$, and $B$  \cite{krzy19, bara19, agbazo21, bara15}. $\alpha_{max}$ and $\alpha_{min}$ denote the most extreme and smoothest event in the considered data-set and $\alpha_{0}$ (=value of $\alpha$ at $f(\alpha)_{max}$) provides the information about the structure of the process, e.g. lower value will signify more correlated process which loose its fine structure and appears to be more regular \cite{agbazo21}. In addition, left skewed multifractal spectrum indicates the dominance of large fluctuations which can lead to extreme events \cite{bara19,bara15}.
\begin{table}[ht]
\centering
\resizebox{0.65\textwidth}{!}{%
\begin{tabular}{|l|l|l|l|l|l|l|l|}
\hline
Sectors & $\Delta \alpha$ & $\alpha_{max}$ & $\alpha_{0}$ & $\alpha_{min}$ & $H_2$    & dH   & B     \\ \hline
AU      & 1.71            & 1.92           & 0.99         & 0.21           & 0.72 & 1.38 & -0.09 \\ \hline
BM      & 0.64            & 0.69           & 0.4          & 0.05           & 0.33 & 0.46 &\ 0.09  \\ \hline
BX      & 1.07            & 1.54           & 0.97         & 0.47           & 0.86 & 0.75 & -0.08 \\ \hline
CD      & 1.21            & 1.37           & 0.59         & 0.16           & 0.47 & 0.93 & -0.29 \\ \hline
CDGS    & 0.46            & 0.83           & 0.73         & 0.37           & 0.66 & 0.3  &\ 0.57  \\ \hline
CG      & 1.55            & 1.51           & 0.75         & -0.04          & 0.5  & 1.15 & -0.02 \\ \hline
CPSE    & 1.4             & 1.61           & 0.92         & 0.21           & 0.71 & 1.09 & \ 0.02  \\ \hline
EG      & 0.58            & 1.07           & 0.82         & 0.49           & 0.82 & 0.35 &\  0.13  \\ \hline
FMCG    & 1.27            & 1.49           & 0.85         & 0.22           & 0.67 & 0.98 &\  0.0   \\ \hline
FN      & 0.71            & 0.86           & 0.54         & 0.15           & 0.44 & 0.55 &\  0.11  \\ \hline
HC      & 0.85            & 1.16           & 0.83         & 0.31           & 0.75 & 0.55 &\  0.23  \\ \hline
ID      & 0.71            & 1.01           & 0.51         & 0.3            & 0.47 & 0.48 & -0.42 \\ \hline
II      & 0.87            & 1.33           & 1            & 0.46           & 0.87 & 0.65 &\  0.26  \\ \hline
IT      & 0.4             & 0.76           & 0.67         & 0.36           & 0.62 & 0.26 &\  0.51  \\ \hline
MT      & 0.89            & 0.59           & 0.32         & -0.29          & 0.18 & 0.65 &\  0.37  \\ \hline
ONG     & 1.04            & 1.23           & 0.52         & 0.19           & 0.45 & 0.68 & -0.36 \\ \hline
PSU     & 1.65            & 1.6            & 0.71         & -0.04          & 0.47 & 1.25 & -0.08 \\ \hline
PWR     & 1.01            & 1.43           & 0.87         & 0.42           & 0.8  & 0.7  & -0.11 \\ \hline
RE      & 0.41            & 0.86           & 0.71         & 0.45           & 0.84 & 0.22 &\  0.23  \\ \hline
TC      & 1.51            & 1.75           & 0.84         & 0.23           & 0.67 & 1.12 & -0.19 \\ \hline
Teck    & 0.82            & 1.2            & 1.01         & 0.38           & 0.84 & 0.63 & \ 0.54  \\ \hline
UT      & 0.66            & 1.2            & 1.03         & 0.54           & 0.91 & 0.44 &\  0.50  \\ \hline
\end{tabular}%
}
\caption{Sector-wise values of width of singularity spectrum ($\Delta \alpha$, column 2), Hurst exponents ($H_2$, column 6), spread in Hurst exponent ($dH$, column 7), and asymmetry ratio ($B$, column 8)}
\label{table2}
\end{table}
A right skewed spectrum, in contrast, signifies presence of more small fluctuations and thus is a sign of stable sector. AU, BX, CG, FMCG, HC, IT, RE, Teck, PWR, CPSE, II, CDGS, EG, TC, and UT sectors with Hurst exponent value greater than $0.5$, indicates the presence of persistent fluctuations or positive auto-correlation [38]. 
On the other hand, CD, MT, ONG, PSU, BM, FN, and ID have anti-persistence behavior or negative correlation. In other words, a positive (negative) return on a specific day will mostly be followed by a negative (positive) return on the next working day. The width of the generalized Hurst exponents $(dH)$ is computed using all $H(q)$ values over the range of $q \epsilon [-10, 10]$. This width signifies the strength of the multifractality in a sector; the higher the values, the higher the multifractality \cite{telesca05}. AU sector $(dH = 1.38)$ is the most multifractal among all the sectors present in Indian market, followed by PSU $(dH = 1.25)$ and CG $(dH = 1.15)$. On the other hand, RE $(dH = 0.22)$, IT $(dH = 0.26)$, CDGS $(dH = 0.3)$ have the lower degree of multifractality in the considered period. This information also gives an idea of long-range dependency \cite{arshad19, rizvi14,anag16,  samadder13, domino11}. We can conclude that RE, IT, and CDGS show the lowest level of dependency among all. Value of $B$ parameter included in Table \ref{table2}  thus helps one to study the dynamics of the market quantitatively, as well as qualitatively \cite{ bara19,bara15}.
\section{Conclusion}\label{sec:conc}
This paper analyzes multifractality of  sectoral daily indices of Bombay stock exchange, India. We replace MFDFA method to OMW-EEMD-MFDFA which makes our method more efficient and analysis robust. Firstly, multifractality in finance is not new, But so far, it has not been analyzed to provide information on sector-wise data in a market. Assessing the economic and financial condition is essential for policymakers, mutual funds, portfolio managers, and investors. It is important to note here that a market's overall efficiency may not always be the same as the efficiency found at the sectoral level. This motivates the authors of these sector-specific studies, also the first comprehensive sector-wise analysis using multifractality. Thus the present work focuses on the sectoral efficiency of $22$ sectors present in the S\&P BSE market in the considered period using multifractality analysis. 

Secondly, this study also proves the existence of long-range correlation and randomness, which depend on the chosen time scales. In this paper, we have figured out the two market dynamics, $10$ to $200$ days as short and more than $200$ days as the long term. It is beneficial for the investors to get the behavior of the individual sectors in the market and how they behave on these two separate scales. Moreover, at shorter time scale, negative $q$ power of variances vanish. At both the scales, decreasing Hurst exponent function with $q$ and non-linear Renyi exponent confirms the presence of multifractality in the data-set. We quantify the multifractality using the width and asymmetry of the multifractal spectrum. The width of the single humped singularity spectrum is more for shorter time period indicating more multifractality for the same. We have separated the sectors in two groups on the basis of persistent and anti-persistent behavior which will help in making investment policies for future.
We also find that AU and RE sectors are the most and least multifractal among all the sectors. This study reveals that the market at a short time scale is highly volatile; hence, short-scale investment (less than $200$ days) is risky. Due to the presence of long-range correlation, long-term trading of more than $200$ days will be a better decision for profitable investment. We hope our study will help the investors keeping in mind that the sectors behave differently comparing each other as well as in different time scales and can be used to make the better strategy by finding the right sector for investment. 

The authors acknowledge the Bombay Stock Exchange (BSE) archive for providing the data \cite{bse}.


 \bibliography{manus_131022}

\begin{thebibliography}{10}
\expandafter\ifx\csname url\endcsname\relax
  \def\url#1{\texttt{#1}}\fi
\expandafter\ifx\csname urlprefix\endcsname\relax\def\urlprefix{URL }\fi
\expandafter\ifx\csname href\endcsname\relax
  \def\href#1#2{#2} \def\path#1{#1}\fi

\bibitem{aslam2020}
F.~Aslam, W.~Mohti, P.~Ferreira, Evidence of intraday multifractality in
  european stock markets during the recent coronavirus (covid-19) outbreak,
  International Journal of Financial Studies 8~(2) (2020) 31.

\bibitem{aslam2021}
F.~Aslam, P.~Ferreira, H.~Ali, S.~Kauser, Herding behavior during the covid-19
  pandemic: A comparison between asian and european stock markets based on
  intraday multifractality, Eurasian Economic Review (2021) 1--27.

\bibitem{zhang2021}
S.~Zhang, W.~Fang, Multifractal behaviors of stock indices and their ability to
  improve forecasting in a volatility clustering period, Entropy 23~(8) (2021)
  1018.

\bibitem{milocs2020}
L.~R. Milo{\c{s}}, C.~Ha{\c{t}}iegan, M.~C. Milo{\c{s}}, F.~M. Barna,
  C.~Boțoc, Multifractal detrended fluctuation analysis (mf-dfa) of stock
  market indexes. empirical evidence from seven central and eastern european
  markets, Sustainability 12~(2) (2020) 535.

\bibitem{kantel02}
J.~W. Kantelhardt, S.~A. Zschiegner, E.~Koscielny-Bunde, S.~Havlin, A.~Bunde,
  H.~E. Stanley, Multifractal detrended fluctuation analysis of nonstationary
  time series, Physica A: Statistical Mechanics and its Applications 316~(1-4)
  (2002) 87--114.

\bibitem{kantel03}
J.~W. Kantelhardt, D.~Rybski, S.~A. Zschiegner, P.~Braun, E.~Koscielny-Bunde,
  V.~Livina, S.~Havlin, A.~Bunde, Multifractality of river runoff and
  precipitation: comparison of fluctuation analysis and wavelet methods,
  Physica A: Statistical Mechanics and its Applications 330~(1-2) (2003)
  240--245.

\bibitem{he17}
S.~He, Y.~Wang, Revisiting the multifractality in stock returns and its
  modeling implications, Physica A: Statistical Mechanics and its Applications
  467 (2017) 11--20.

\bibitem{sim19}
S.~Lai, L.~Wan, X.~Zeng, Comparative study of sliding window multifractal
  detrended fluctuation analysis and multifractal moving average algorithm, in:
  Journal of Physics: Conference Series, Vol. 1345, IOP Publishing, 2019, p.
  042086.

\bibitem{zho12}
W.-X. Zhou, Finite-size effect and the components of multifractality in
  financial volatility, Chaos, Solitons \& Fractals 45~(2) (2012) 147--155.

\bibitem{sus20}
L.~R. Milo{\c{s}}, C.~Ha{\c{t}}iegan, M.~C. Milo{\c{s}}, F.~M. Barna,
  C.~Boțoc, Multifractal detrended fluctuation analysis (mf-dfa) of stock
  market indexes. empirical evidence from seven central and eastern european
  markets, Sustainability 12~(2) (2020) 535.

\bibitem{charutha2020}
S.~Charutha, M.~G. Krishna, P.~Manimaran, Multifractal analysis of indian
  public sector enterprises, Physica A: Statistical Mechanics and its
  Applications 557 (2020) 124881.

\bibitem{lai19}
S.~Lai, L.~Wan, X.~Zeng, Comparative study of sliding window multifractal
  detrended fluctuation analysis and multifractal moving average algorithm, in:
  Journal of Physics: Conference Series, Vol. 1345, IOP Publishing, 2019, p.
  042086.

\bibitem{arshad19}
S.~Arshad, S.~A.~R. Rizvi, O.~Haroon, Understanding asian emerging stock
  markets, Bulletin of Monetary Economics and Banking 21 (2019) 495--510.

\bibitem{gu2013}
R.~Gu, Y.~Shao, Q.~Wang, Is the efficiency of stock market correlated with
  multifractality? an evidence from the shanghai stock market, Physica A:
  Statistical Mechanics and its Applications 392~(2) (2013) 361--370.

\bibitem{dutta2010}
S.~Dutta, Multifractal detrended fluctuation analysis of sensex fluctuation in
  the indian stock market, Canadian Journal of Physics 88~(8) (2010) 545--551.

\bibitem{hasan15}
R.~Hasan, S.~M. Mohammad, Multifractal analysis of asian markets during
  2007--2008 financial crisis, Physica A: Statistical Mechanics and its
  Applications 419 (2015) 746--761.

\bibitem{caraiani2012evidence}
P.~Caraiani, Evidence of multifractality from emerging european stock markets,
  PloS one 7~(7) (2012) e40693.

\bibitem{grech2009}
D.~Grech, L.~Czarnecki, Multifractal dynamics of stock markets, arXiv preprint
  arXiv:0912.3390 (2009).

\bibitem{tiwari2019}
A.~K. Tiwari, G.~C. Aye, R.~Gupta, Stock market efficiency analysis using long
  spans of data: A multifractal detrended fluctuation approach, Finance
  Research Letters 28 (2019) 398--411.

\bibitem{rizvi14}
S.~A.~R. Rizvi, S.~Arshad, Investigating the efficiency of east asian stock
  markets through booms and busts, Pacific Science Review 16~(4) (2014)
  275--279.

\bibitem{anag16}
P.~Anagnostidis, C.~Varsakelis, C.~J. Emmanouilides, Has the 2008 financial
  crisis affected stock market efficiency? the case of eurozone, Physica A:
  statistical mechanics and its applications 447 (2016) 116--128.

\bibitem{kumar2009}
S.~Kumar, N.~Deo, Multifractal properties of the indian financial market,
  Physica A: Statistical Mechanics and its Applications 388~(8) (2009)
  1593--1602.

\bibitem{nargunam2022}
R.~Nargunam, A.~Lahiri, Persistence in daily returns of stocks with highest
  market capitalization in the indian market, Digital Finance (2022) 1--34.

\bibitem{zhou10}
Y.~Zhou, Y.~Leung, Multifractal temporally weighted detrended fluctuation
  analysis and its application in the analysis of scaling behavior in
  temperature series, Journal of Statistical Mechanics: Theory and Experiment
  2010~(06) (2010) P06021.

\bibitem{qian11}
X.-Y. Qian, G.-F. Gu, W.-X. Zhou, Modified detrended fluctuation analysis based
  on empirical mode decomposition for the characterization of anti-persistent
  processes, Physica A: Statistical Mechanics and its Applications 390~(23-24)
  (2011) 4388--4395.

\bibitem{wu09}
Z.~Wu, N.~E. Huang, Ensemble empirical mode decomposition: a noise-assisted
  data analysis method, Advances in adaptive data analysis 1~(01) (2009) 1--41.

\bibitem{lee18}
M.~Lee, J.~W. Song, S.~Kim, W.~Chang, Asymmetric market efficiency using the
  index-based asymmetric-mfdfa, Physica A: Statistical Mechanics and Its
  Applications 512 (2018) 1278--1294.

\bibitem{bse}
Bse india database, \url{https://www.bseindia.com/}, accessed on 19th July,
  2022, using the programming language and numeric computing environment
  developed by MathWorks. (2017).

\bibitem{gan15}
E.~Li, X.~Mu, G.~Zhao, P.~Gao, Multifractal detrended fluctuation analysis of
  streamflow in the yellow river basin, china, Water 7~(4) (2015) 1670--1686.

\bibitem{krzy19}
J.~Krzyszczak, P.~Baranowski, M.~Zubik, V.~Kazandjiev, V.~Georgieva,
  C.~S{\l}awi{\'n}ski, K.~Siwek, J.~Kozyra, A.~Nier{\'o}bca, Multifractal
  characterization and comparison of meteorological time series from two
  climatic zones, Theoretical and Applied Climatology 137~(3) (2019)
  1811--1824.

\bibitem{adarsh19}
S.~Adarsh, D.~N. Kumar, B.~Deepthi, G.~Gayathri, S.~Aswathy, S.~Bhagyasree,
  Multifractal characterization of meteorological drought in india using
  detrended fluctuation analysis, International Journal of Climatology 39~(11)
  (2019) 4234--4255.

\bibitem{bara19}
P.~Baranowski, M.~Gos, J.~Krzyszczak, K.~Siwek, A.~Kieliszek, P.~Tkaczyk,
  Multifractality of meteorological time series for poland on the base of
  merra-2 data, Chaos, Solitons \& Fractals 127 (2019) 318--333.

\bibitem{peng94}
C.-K. Peng, S.~V. Buldyrev, S.~Havlin, M.~Simons, H.~E. Stanley, A.~L.
  Goldberger, Mosaic organization of dna nucleotides, Physical review e 49~(2)
  (1994) 1685.

\bibitem{mandel97}
B.~B. Mandelbrot, A.~J. Fisher, L.~E. Calvet, A multifractal model of asset
  returns (1997).

\bibitem{movahed06}
M.~S. Movahed, G.~Jafari, F.~Ghasemi, S.~Rahvar, M.~R.~R. Tabar, Multifractal
  detrended fluctuation analysis of sunspot time series, Journal of Statistical
  Mechanics: Theory and Experiment 2006~(02) (2006) P02003.

\bibitem{ihlen12}
E.~A. Ihlen, Introduction to multifractal detrended fluctuation analysis in
  matlab, Frontiers in physiology 3 (2012) 141.

\bibitem{wu07}
Z.~Wu, N.~E. Huang, S.~R. Long, C.-K. Peng, On the trend, detrending, and
  variability of nonlinear and nonstationary time series, Proceedings of the
  National Academy of Sciences 104~(38) (2007) 14889--14894.

\bibitem{zhang19}
X.~Zhang, G.~Zhang, L.~Qiu, B.~Zhang, Y.~Sun, Z.~Gui, Q.~Zhang, A modified
  multifractal detrended fluctuation analysis (mfdfa) approach for multifractal
  analysis of precipitation in dongting lake basin, china, Water 11~(5) (2019)
  891.

\bibitem{liu20}
X.~Liu, C.~Xia, Z.~Chen, Y.~Chai, R.~Jia, A new framework for rainfall
  downscaling based on eemd and an improved fractal interpolation algorithm,
  Stochastic Environmental Research and Risk Assessment 34~(8) (2020)
  1147--1173.

\bibitem{huang08}
N.~E. Huang, Z.~Wu, A review on hilbert-huang transform: Method and its
  applications to geophysical studies, Reviews of geophysics 46~(2) (2008).

\bibitem{agbazo21}
M.~N. Agbazo, G.~K. N’Gobi, E.~Alamou, B.~Kounouhewa, A.~Afouda, Assessing
  nonlinear dynamics and trends in precipitation by ensemble empirical mode
  decomposition (eemd) and fractal approach in benin republic (west africa),
  Complexity 2021 (2021).

\bibitem{peng05}
Z.~Peng, W.~T. Peter, F.~Chu, An improved hilbert--huang transform and its
  application in vibration signal analysis, Journal of sound and vibration
  286~(1-2) (2005) 187--205.

\bibitem{ayenu10}
A.~Ayenu-Prah, N.~Attoh-Okine, A criterion for selecting relevant intrinsic
  mode functions in empirical mode decomposition, Advances in Adaptive Data
  Analysis 2~(01) (2010) 1--24.

\bibitem{hurst51}
H.~E. Hurst, Long-term storage capacity of reservoirs, Transactions of the
  American society of civil engineers 116~(1) (1951) 770--799.

\bibitem{hentschel83}
H.~G.~E. Hentschel, I.~Procaccia, The infinite number of generalized dimensions
  of fractals and strange attractors, Physica D: Nonlinear Phenomena 8~(3)
  (1983) 435--444.

\bibitem{halsey86}
T.~C. Halsey, M.~H. Jensen, L.~P. Kadanoff, I.~Procaccia, B.~I. Shraiman,
  Fractal measures and their singularities: The characterization of strange
  sets, Physical review A 33~(2) (1986) 1141.

\bibitem{kurths87}
J.~Kurths, H.~Herzel, An attractor in a solar time series, Physica D: Nonlinear
  Phenomena 25~(1-3) (1987) 165--172.

\bibitem{meneveau87}
C.~Meneveau, K.~R. Sreenivasan, The multifractal spectrum of the dissipation
  field in turbulent flows, Nuclear Physics B-Proceedings Supplements 2 (1987)
  49--76.

\bibitem{salat17}
H.~Salat, R.~Murcio, E.~Arcaute, Multifractal methodology, Physica A:
  Statistical Mechanics and its Applications 473 (2017) 467--487.

\bibitem{weron04}
R.~Weron, I.~Simonsen, P.~Wilman, Modeling highly volatile and seasonal
  markets: evidence from the nord pool electricity market, in: The application
  of econophysics, Springer, 2004, pp. 182--191.

\bibitem{simonsen03}
I.~Simonsen, Measuring anti-correlations in the nordic electricity spot market
  by wavelets, Physica A: Statistical Mechanics and its applications 322 (2003)
  597--606.

\bibitem{gorjao21}
L.~R. Gorj{\~a}o, D.~Witthaut, P.~G. Lind, W.~Medjroubi, Change of persistence
  in european electricity spot prices, arXiv preprint arXiv:2112.03513 (2021).

\bibitem{gorjao22}
L.~R. Gorj{\~a}o, G.~Hassan, J.~Kurths, D.~Witthaut, Mfdfa: Efficient
  multifractal detrended fluctuation analysis in python, Computer Physics
  Communications 273 (2022) 108254.

\bibitem{han22}
C.~Han, H.~Hilger, E.~Mix, P.~C. B{\"o}ttcher, M.~Reyers, C.~Beck, D.~Witthaut,
  L.~R. Gorj{\~a}o, Complexity and persistence of price time series of the
  european electricity spot market, PRX Energy 1~(1) (2022) 013002.

\bibitem{zunino08}
L.~Zunino, B.~M. Tabak, A.~Figliola, D.~G. P{\'e}rez, M.~Garavaglia, O.~A.
  Rosso, A multifractal approach for stock market inefficiency, Physica A:
  Statistical Mechanics and its Applications 387~(26) (2008) 6558--6566.

\bibitem{agter01}
F.~P. Agterberg, Multifractal simulation of geochemical map patterns, in:
  Geologic modeling and simulation, Springer, 2001, pp. 327--346.

\bibitem{hou18}
W.~Hou, G.~Feng, P.~Yan, S.~Li, Multifractal analysis of the drought area in
  seven large regions of china from 1961 to 2012, Meteorology and Atmospheric
  Physics 130~(4) (2018) 459--471.

\bibitem{xie04}
S.~Xie, Z.~Bao, Fractal and multifractal properties of geochemical fields,
  Mathematical Geology 36~(7) (2004) 847--864.

\bibitem{bara15}
P.~Baranowski, J.~Krzyszczak, C.~Slawinski, H.~Hoffmann, J.~Kozyra,
  A.~Nier{\'o}bca, K.~Siwek, A.~Gluza, Multifractal analysis of meteorological
  time series to assess climate impacts, Climate Research 65 (2015) 39--52.

\bibitem{telesca05}
L.~Telesca, V.~Lapenna, M.~Macchiato, Multifractal fluctuations in seismic
  interspike series, Physica A: Statistical Mechanics and its Applications 354
  (2005) 629--640.

\bibitem{samadder13}
S.~Samadder, K.~Ghosh, T.~Basu, Fractal analysis of prime indian stock market
  indices, Fractals 21~(01) (2013) 1350003.

\bibitem{domino11}
K.~Domino, The use of the hurst exponent to predict changes in trends on the
  warsaw stock exchange, Physica A: Statistical Mechanics and its Applications
  390~(1) (2011) 98--109.

\end{thebibliography}

\end{document}